# Dressed-state control of effective dipolar interaction between strongly-coupled solid-state spins

**Authors:** Junghyun Lee[1,2], Mamiko Tatsuta[3], Andrew Xu[1,*], Erik Bauch[4], Mark J. H. Ku[4,5,†], and Ronald L. Walsworth[4,5,6,‡]

**Affiliations:**

[1]Department of Physics, Massachusetts Institute of Technology, Cambridge, Massachusetts 02139, USA.

[2]Center for Quantum Information, Korea Institute of Science and Technology, Seoul 02792, Republic of Korea.

[3]Research Center for Emerging Computing Technologies, National institute of Advanced Industrial Science and Technology, Tsukuba, Ibaraki 305-8568, Japan.

[4]Harvard-Smithsonian Center for Astrophysics, Cambridge, Massachusetts 02138, USA.

[5]Department of Physics, Harvard University, Cambridge, Massachusetts 02138, USA.

[6]Center for Brain Science, Harvard University, Cambridge, Massachusetts 02138, USA.

*Correspondence to: walsworth@umd.edu

\* Now at Pritzker School of Molecular Engineering, University of Chicago, Chicago, IL 60637, USA

† Now at Department of Physics and Astronomy, University of Delaware

‡ Now at Department of Physics, University of Maryland

**Abstract:**

Strong interactions between spins in many-body solid-state quantum systems are a crucial resource for exploring and applying non-classical states. In particular, electronic spins associated with defects in diamond system are a leading platform for the study of collective quantum phenomena and for quantum technology applications. While such solid-state quantum defect systems have the advantage of spin ensemble control and operation under ambient conditions, they face the key challenge of controlling interactions between the defects spins, since the defects are spatially fixed inside the host lattice with relative positions that cannot be well controlled during fabrication. In this work, we present a dressed-state approach to control the effective dipolar coupling between solid-state spins; and then demonstrate this scheme experimentally using two strongly-coupled nitrogen vacancy (NV) centres in diamond. Including Rabi driving terms between the $m_s = 0$ and $\pm 1$ states in the NV spin Hamiltonian allows us to turn on and off or tune the effective dipolar coupling between two NV spins, $NV_A$ and $NV_B$. Through Ramsey spectroscopy on the probe $NV_A$ spin, we detect the change of the effective dipolar field generated by the control $NV_B$ spin prepared in different dressed states. To observe the change of interaction dynamics, we then deploy spin-lock-based polarization transfer measurements via a Hartmann-Hahn matching condition between the two NV spins in different dressed states. We also perform simulations that indicate the promise for this scheme to control the distribution of interaction strengths in strongly interacting spin systems, which can be a valuable tool for studying non-equilibrium quantum phases and generating high fidelity multi-spin correlated states for quantum-enhanced sensing.

**Main text:**

**Introduction**

Understanding and engineering of strongly-coupled solid-state quantum spin systems is a key challenge for quantum technology. Such systems could be utilized to observe and generate collective quantum behaviour, leading to highly sought-after applications ranging from quantum simulation of non-equilibrium phases[1–3] to quantum enhanced sensing applications beyond the classical limit[4–6]. In particular, recent work using nitrogen-vacancy (NV) centres in diamond have addressed challenging problems such as the observation of critical thermalization in a three dimensional ensemble[1], and a Discrete Time Crystal (DTC) state subject to a periodic drive in a disordered spin ensemble[2]. Realization of a strongly interacting, many-spin system was possible through the fabrication of dense NV ensembles in diamond samples with both large nitrogen density ([N]) and high [N] to [NV] conversion yield[1,7]. In this regime, NV-NV dipolar couplings are the dominant spin interactions[1,8], with robust control possible for an ensemble of NV spins at ambient temperature.

To build on this progress and realize the aforementioned applications, it is crucial to have deterministic control of interactions between the strongly-coupled solid-state spins. To date, such control has not been possible in a scalable manner, due to variation in spin-spin separation from the stochastic process by which quantum defects are fabricated in the host solid[9]. Nanoscale spatial precision of defect formation has recently been demonstrated[10, 11], yet the generation of ensembles of solid-state spins still largely depends on the stochastic methods of ion implantation[12, 13] and chemical vapor deposition[14], leading to an order of magnitude variation in the distribution of spin-spin interactions. The lack of control over such spin-spin interactions limits the utility of solid-state spin systems for many-body simulations and the generation of multipartite entanglement for quantum-enhanced sensing[15].

In this work, we present a method to deterministically tune the dipolar coupling between strongly-coupled spin-1 qutrits through the manipulation of dressed states. We use a strongly-coupled pair of NV centres in diamond as a testbed to demonstrate the approach. The negatively charged NV centre is effectively a two-electron system[16], forming an S = 1 spin qutrit. The present study builds on past work where pairs of strongly interacting electronic spins in diamond were utilized to study coherent manipulation of an electronic dark spin[17], as well as generation of a room temperature entangled state[18]. Here, we employ two Rabi driving fields to induce an oscillating NV spin population between both the $m_s = 0$ and $-1$ level and the $m_s = 0$ and $+1$ level, thereby generating a spin qutrit dressed states. With such a doubly dressed states, an effective dipolar coupling between the two NV spins (labelled $NV_A$ and $NV_B$) can be modulated by careful apportion of the relative Rabi driving field magnitudes. In future work, this technique may enable robust tuning of interactions within an NV ensemble spin system, allowing quantum enhanced sensing, and engineering of the NV-NV coupling dynamics or local disorder amplitude to study transitions of non-equilibrium phases[1, 2].

Generating dressed states in an interacting spin-1/2 system via introducing driving fields to decouple dipolar interactions (also referred to as motional narrowing) has been extensively studied in diverse systems such as NMR[19, 20] and superconducting qubits[21]. For spin defects in diamond, a single Rabi driving field was applied to spin-1/2 nitrogen electronic spins (P1s) to suppress the overall dipolar field noise on NV spins[22, 23]. However, more complicated dynamics can arise for strongly-coupled spin-1 qutrits[24], thereby requiring further analysis and experimental investigation, as presented here.

To begin, we consider a simple two qutrit spin Hamiltonian with driving terms. With the spin flip-flop terms neglected for the dipolar interaction between the NV electronic spins[1], the effective time-dependent Hamiltonian of the system with resonant driving fields is given by:

$$\hat{H}(t) = D(S_A^z)^2 + \gamma B_A S_A^z + D(S_B^z)^2 + \gamma B_B S_B^z \\ + \sum_{i \in \{+,-\}} \Omega_i^A \cos(\omega_i^A t) S_{A,i}^x + \sum_{i \in \{+,-\}} \Omega_i^B \cos(\omega_i^B t) S_{B,i}^x + \nu_{dip} S_A^z S_B^z \quad (1)$$

where $D$ denotes the zero-field splitting; $\vec{S}_A$ and $\vec{S}_B$ are the NV electronic spin operators; $\gamma B_i$ is the bias magnetic field Zeeman splitting for each NV spin with field projection $B_i$ on each NV and gyromagnetic ratio $\gamma$; $\nu_{dip}$ parameterizes the magnetic dipolar coupling between the two NV spins; and $\Omega_\pm^{A,B}$, $\omega_\pm^{A,B}$ are the Rabi and carrier frequencies, respectively, of external microwave fields that resonantly drive the $|m_s = 0\rangle \leftrightarrow |m_s = +1\rangle$ and $|m_s = 0\rangle \leftrightarrow |m_s = -1\rangle$ transitions for NV$_A$ and NV$_B$. Hyperfine and transverse Zeeman terms are neglected in the above Hamiltonian, as their effect is greatly suppressed given the large NV zero-field splitting ($D$ = 2.87 GHz), as well as the bias magnetic field and Rabi drive frequency conditions used in the present experiment. One exception is energy-level splitting of the NV spin sublevels due to the non-secular hyperfine interaction, as observed in Ramsey measurements and discussed below. (See additional discussion in supplementary material section 1.)

In our experiment, we apply Rabi driving to the "control" spin NV$_B$ while detecting the resulting effect with the second "sensor" spin NV$_A$, with the bias magnetic field aligned with NV$_A$. By taking $\Omega_{+,-}^A = 0$ and $\Omega_{+,-}^B = \Omega_{+,-}$, the overall Hamiltonian can be diagonalized in a doubly rotating frame with the rotating wave approximation. The effective dipolar coupling term can then be analytically solved, under the condition of $\nu_{dip} \ll \Omega_{+,-}$, yielding (see supplementary material section 1.2):

$$\nu_{eff} \approx \frac{1}{2} \frac{(\Omega_+^2 - \Omega_-^2)}{(\Omega_+^2 + \Omega_-^2)} \nu_{dip} \quad (2)$$

The resulting doubly dressed states can be projected onto an effective spin-1/2 system, as follows:

$$|+\rangle_d = \frac{1}{\sqrt{2}}\left(\frac{\Omega_+}{\sqrt{\Omega_+^2 + \Omega_-^2}}|+\rangle + \frac{\Omega_-}{\sqrt{\Omega_+^2 + \Omega_-^2}}|-\rangle + |0\rangle\right),$$

$$|-\rangle_d = \frac{1}{\sqrt{2}}\left(\frac{\Omega_+}{\sqrt{\Omega_+^2 + \Omega_-^2}}|+\rangle + \frac{\Omega_-}{\sqrt{\Omega_+^2 + \Omega_-^2}}|-\rangle - |0\rangle\right) \quad (3)$$

In this framework, we define new eigenvectors, denoted as $\frac{\Omega_+}{\sqrt{\Omega_+^2+\Omega_-^2}}|+\rangle + \frac{\Omega_-}{\sqrt{\Omega_+^2+\Omega_-^2}}|-\rangle$ and $|0\rangle$, for the rotated Pauli operator $S_R^z$. The effective spin coupling can be interpreted as the projection of the lab frame interaction component, $\nu_{dip} S_A^z \otimes S_B^z$, onto $\nu_{eff} S_A^z \otimes \left(S_{B,R}^z\right)^2$, where $\nu_{dip}$ is transformed to $\nu_{eff}$ due to the projection factor $\frac{1}{2}\frac{(\Omega_+^2-\Omega_-^2)}{(\Omega_+^2+\Omega_-^2)}$. Eq. (2) indicates that by tuning $\Omega_{+,-}$, we can deterministically vary the effective dipolar coupling strength between $-\nu_{dip}/2$ and $+\nu_{dip}/2$. Eq. (2) is a generalized formula that provides different driving conditions; e.g., $\Omega_+ = \Omega_-, \nu_{eff} = 0; \Omega_- = 0, \nu_{eff} = \frac{\nu_{dip}}{2}$; and $\Omega_+ = 0, \nu_{eff} = -\nu_{dip}/2$.

**Results**

To demonstrate the tunable effective coupling via a doubly dressed state scheme within the solid-state spins, we use two strongly-coupled NV qutrit spins as a simple model system (with $\nu_{dip} > \Delta_{bath} \approx 1/T_2^*$, where $\Delta_{bath}$ is the effective coupling strength between the NV and bath spins). To realize an isolated two strongly-coupled NV spins, we use a molecular implantation technique[17,25]. More details on the diamond sample is discussed in the methods. We first deploy a double electron-electron resonance (DEER) measurement protocol to measure the intrinsic coupling strength between two neighboring and strongly-coupled NV spins. Due to DEER's spin echo based pulse scheme, the NV$_A$ sensor spin accumulates net phase only due to the repeated inversion of the NV$_B$ control spin, filtering out other possible magnetic signal sources at frequencies lower than the NV$_B$ spin modulation. We then perform Ramsey spectroscopy on NV$_A$ for different polarizations and dressed states of NV$_B$: this measurement characterizes the effective NV-NV dipolar field under different dressed states. To directly observe the change of NV-NV interaction dynamics under different dressed states, NV-NV polarization transfer via a spin-lock pulse sequence is used (Fig. 4). Note that the roles of NV$_A$ and NV$_B$ can be interchanged in all the results presented here. Finally, we employ Monte Carlo simulations to estimate variations in the interaction dynamics of the ensemble spin system utilizing the doubly dressed state method.

## A. Measuring the dipolar coupling strength using DEER

We deploy a DEER pulse sequence to determine the strength of the bare dipolar coupling between two strongly interacting NV spins. We select neighboring NVs with different quantization axes (Fig. 1a), which allows us to distinguish the two NV spins in the ODMR frequency domain (Fig. 1b), thereby allowing individual NV spin control. The DEER technique measures the dynamic phase accumulated by the sensor spin (NV$_A$) due to the dipolar magnetic field generated by the control spin (NV$_B$). First, we performed a DEER measurement in the single quantum (SQ) basis of $|0\rangle$ and $|+1\rangle$ states. The NV$_A$ and NV$_B$ interaction is given by an Ising term with coupling strength $\nu_{dip}$ (Equation (1)). The accumulated phase is projected back to the $|0\rangle$ spin state for NV$_A$ via a probability measurement $P_{SQ} \propto cos(\nu_{dip}\tau/2)$. Meanwhile, under the double quantum (DQ) basis of $|B\rangle = \frac{|+1\rangle+|-1\rangle}{\sqrt{2}}$ and $|D\rangle = \frac{|+1\rangle-|-1\rangle}{\sqrt{2}}$ for NV$_A$, and $|-1\rangle, |+1\rangle$ states for NV$_B$, the accumulated phase projected back to $|0\rangle$ for NV$_A$ is given by $P_{DQ} \propto cos(2\nu_{dip}\tau)$. In the SQ basis of sensor spin NV$_A$ with the control spin NV$_B$ flipped to $|+1\rangle$ after being initialized to $|0\rangle$, we measure a DEER signal oscillation of $\nu_{dip}/2 = 0.125 \pm 0.01$ MHz (see Fig. 1c, left). In the DQ basis of sensor spin NV$_A$ with the control spin NV$_B$ flipped to $|+1\rangle$ after being initialized to $|-1\rangle$, we measure a DEER signal oscillation of $2\nu_{dip} = 0.495 \pm 0.031$ MHz (see Fig. 1c, right). From both these measurements, we extract the NV$_A$-NV$_B$ dipolar coupling parameter $\nu_{dip} = 0.250 \pm 0.015$ MHz.

## B. Measuring the dipolar coupling strength using Ramsey interferometry

Similar to DEER, Ramsey interferometry also allows the sensor spin NV$_A$ to accumulate a dynamic phase due to the static dipolar field produced by the control spin NV$_B$. We can further extend the Ramsey method to measure the change in effective coupling strength by transforming NV$_B$ into different dressed states. In past work, Ramsey spectroscopy has been used as a spectroscopic tool to measure the effective coupling strengths between an ensemble of NV spins and a bath of paramagnetic P1 spins[26]. In contrast to DEER, there is no $\pi$ pulse applied to NV$_B$ during a Ramsey measurement; therefore, the dipolar field is constant for an initially prepared $m_s$ state of NV$_B$ during the phase accumulation of the NV$_A$ spin. For a Ramsey measurement on the NV$_A$ SQ basis of $|0\rangle$ and $|+1\rangle$, when NV$_B$ is prepared in $|0\rangle$, no dynamic phase is accumulated on NV$_A$ due to the zero longitudinal dipolar coupling. However, for NV$_B$ prepared in $|\pm 1\rangle$ with non-zero longitudinal dipolar coupling, NV$_A$ exhibits a phase modulation of $\pm \gamma \nu_{dip} \tau$ during a Ramsey measurement. When a similar Ramsey measurement is performed using the NV$_A$ DQ basis of $|B\rangle$ and $|D\rangle$, the effective magnetic moment of the NV$_A$ spin is doubled and there is thus a twice faster phase accumulation. A fast Fourier transformation (FFT) applied to the Ramsey signal then reveals the phase modulation frequency and hence the dipolar coupling magnitude between the two spins. Our NV spin is accompanied by a $^{14}$N (I = 1) host nuclear spin; and throughout the measurement analysis, we focus on a spin-state projection measurement on the $m_I = 0$ state, one of the three NV hyperfine peaks in the ODMR spectrum. In the NV$_A$ SQ basis, we find Ramsey resonance peak shifts of $\pm \nu_{dip}$ = 0.26 ± 0.02 MHz; see Fig. 2a. In the DQ basis, we determine peak shifts of $\pm 2\nu_{dip}$ = 0.52 ± 0.02 MHz, relative to the Ramsey resonance peak for the non-interacting case; see Fig. 2b. Uncertainty here is given by the frequency resolution of the FFT. Note that the dipolar coupling strength extracted from SQ and DQ Ramsey spectroscopy agrees with the DEER measurements described above. This coupling strength implies a separation of the two NVs of about 6 nm.

**C. Doubly dressed-state control of effective dipolar coupling detected via Ramsey interferometry**

We next introduce a driving field on the control spin NV$_B$ to generate a doubly dressed state, as outlined above; and then perform Ramsey detection in the DQ basis of NV$_A$ to sense the resulting interaction dynamics. In a semi-classical spin picture, double driving of the NV$_B$ spin transitions with Rabi frequencies of $\Omega_+$ and $\Omega_-$ transfer population into each spin state $|+1\rangle$ and $|-1\rangle$ proportional to $\Omega_+^2$ and $\Omega_-^2$, respectively. In the fast driving limit of $\nu_{dip} \ll \Omega_\pm$, each population is time-averaged to its half, and the overall net spin population becomes $\frac{1}{2}(\Omega_+^2 - \Omega_-^2)$. Normalizing to the total population $\Omega_+^2 + \Omega_-^2$, we get a time-averaged effective spin number of $m_s^{eff} \approx (\Omega_+^2 - \Omega_-^2)/2(\Omega_+^2 + \Omega_-^2)$; hence, the effective coupling between NV$_A$ and NV$_B$ becomes $v_{eff} = m_s^{eff} v_{dip}$, which is in agreement with Eqn. (2). In the simplified Hamiltonian framework (see supplementary material section 1.3), the doubly dressed NV$_B$ can be regarded as an effective spin-1/2 system with the dressed states denoted as $|+\rangle_d, |-\rangle_d$; the coupling strength between NV$_A$ and NV$_B$ is given by Eqn (2), which includes the time-averaging effect. See Fig. 3a.

For detection, we monitor the DQ Ramsey power spectrum of one of the NV$_A$ hyperfine peaks in the frequency domain as we change $\Omega_+$ and $\Omega_-$. To satisfy the $v_{dip} \ll \Omega_\pm$ condition, all measurements are done with 2 MHz $< \Omega_\pm$. We define $\alpha = \frac{(\Omega_+ - \Omega_-)}{(\Omega_+ + \Omega_-)}$ as a control parameter that indicates the degree of relative driving strengths. For example, $\alpha = 0$ for equal Rabi frequencies ($\Omega_+ = \Omega_-$), and $\alpha = +1$ when driving only the single transition $|0\rangle \leftrightarrow |+1\rangle$ ($\Omega_+ \neq 0$, $\Omega_- = 0$). As $\alpha$ is swept from $-1$ to $+1$, the Ramsey power spectrum peak transitions from $v_{eff} = +v_{dip} = +0.26$ MHz to $v_{eff} = -v_{dip} = -0.26$ MHz (Fig. 3b). Note that the twice larger variation in measured $v_{eff}$, in contrast to Eqn. (2), is due to the Ramsey spectrum being measured in the DQ basis. We confirm that the effective coupling extracted from Eqn. (2) and our numerical simulation, based on Eqn. (1), lie within the measurement error bound (Fig. 3c). The detuning between each NV spin's resonance frequency $\approx 60$ MHz; therefore, the cross-talk effect from NV$_B$'s driving field on NV$_A$ is negligible (see further discussion below).

### D. Polarization transfer in dressed states

To validate the applicability of the doubly dressed state scheme for controlling spin-spin interactions deterministically, we employ a dressed state spin polarization measurement[27] with a spin-lock (SL) pulse sequence by varying the $|\pm 1\rangle$ transitions relative Rabi frequencies. Polarization transfer measurements are a useful tool for exploring interaction dynamics within strongly-coupled systems[1]. In a two-spin-1/2 system, the cross-polarization transfer rate is given by half the bare dipolar coupling strength when the Rabi frequencies are matched. For the doubly dressed state in a two-spin qutrit system, the effective coupling strength can be tuned within the effective spin-1/2 framework (Fig. 4a) via the interaction term $\frac{v_{eff}}{2} S_A^x \otimes S_B^x$, which implies that one can control the transfer rate deterministically (see supplementary material Eqn. (S29)).

We choose NV$_A$ as the polarization delivering spin and NV$_B$ as the polarization target spin, and observe polarization transfer within the $|0\rangle \leftrightarrow |+1\rangle$ subspace. NV$_B$ is first initialized and prepared in a fully dephased state in the $|0\rangle, |+1\rangle$ basis by applying a $\left(\frac{\pi}{2}\right)_x$ pulse and a wait time $T_{wait} \gg T_2^*$. The purpose of using a dephased NV$_B$ as a target polarization transfer state was to showcase the potential of off-axis NVs as a tunable source of spin bath noise in ensemble spins for many-body simulations, utilizing the double driving scheme introduced in this study. Then NV$_B$ is driven with single or double transition driving fields by tuning $\Omega_+^B(|0\rangle \leftrightarrow |+1\rangle)$ and $\Omega_-^B(|0\rangle \leftrightarrow |-1\rangle)$, inducing a change in the effective dipolar coupling between the two NVs in a dressed state picture (Fig. 4b). NV$_A$ is initialized and spin-locked along the y-axis with driving field $\Omega^A(|0\rangle \leftrightarrow |+1\rangle)$. Once the two NVs satisfy either the singly or doubly dressed state Hartmann-Hahn matching conditions (SHH or DHH)[28], only the $\frac{\nu_{eff}}{2} S_A^x \otimes S_B^x$ term (see supplementary material Eqn. (S29)) survives in the rotating frame Hamiltonian, inducing transfer of polarization from NV$_A$ to NV$_B$. For a two-level spin picture, this can be understood as generating resonant energy levels between the two dressed-state spin qubits. The Rabi frequency for each spin corresponds to the energy level splitting in the rotating frame, and by tailoring the Rabi frequency matched dressed states between the two spins, polarization can be exchanged in the double-rotating frame. For an S = 1 spin system, a doubly dressed state generates an effective two-level system with energy splitting of $\sqrt{\Omega_+^2 + \Omega_-^2}$. Therefore, energy-conserving polarization exchange can occur once the singly driven qutrit spin's Rabi frequency matches the other qutrit spin's doubly driven Rabi frequency $\sqrt{\Omega_+^2 + \Omega_-^2}$. For a given NV$_B$ singly or doubly dressed state, the Rabi frequency of NV$_A$ and the SL duration determine the degree and rate of polarization loss(gain) of NV$_A$ (NV$_B$).

With a fixed spin-lock time $\tau$, Rabi frequency $\Omega^A$ is swept to verify the Hartmann-Hahn (HH) matching condition (see supplementary material Eqn. (S29)) for the dressed state scheme. We apply a $|0\rangle \leftrightarrow |+1\rangle$ single transition driving field to NV$_B$ with Rabi frequency of $\Omega_+^B = 7.56$ MHz, and sweep the NV$_A$ Rabi frequency between $|0\rangle \leftrightarrow |+1\rangle$ from $\Omega^A = 6$ to 9 MHz. Here, the SL duration is set to be the inverse of the estimated effective dipolar coupling between the two NVs. A Lorentzian dip is observed in the NV$_A$ SL coherence measurement (Fig. 4c), indicating a loss of polarization from NV$_A$; the dip is located at $\Omega^A = 7.66 \pm 0.1$ MHz, which coincides with the expected SHH matching condition. Next, we apply a double driving field to NV$_B$, with Rabi frequencies of $\Omega_+^B = 9.59$ MHz and $\Omega_-^B = 4.13$ MHz, to induce a change in polarization transfer dynamics. Again, the NV$_A$ $|0\rangle \leftrightarrow |+1\rangle$ transition Rabi frequency is swept from $\Omega^A = 9$ to 12 MHz. The DHH matching condition is given by $\Omega^A = \sqrt{(\Omega_+^B)^2 + (\Omega_-^B)^2}$ and the measurement result shows a dip appearing at $\Omega^A = 10.51 \pm 0.1$ MHz (Fig. 4c). This result matches well with the calculated DHH condition of $\Omega^A = 10.44$ MHz. Broadening of the DHH polarization transfer dip, compared to that of SHH, may be due to heating of the coplanar waveguide used to deliver microwave signals. Note that the Rabi drive cross-talk error between the NV transitions is on the order of $O\left(\frac{\Omega^2}{\Delta_{A,B}^2}\right)$. For our experiments, $\Delta_{A,B} \approx 60$ MHz is the frequency detuning between the two NV $|0\rangle \leftrightarrow |+1\rangle$ resonance peaks; hence the cross-talk error $\approx 0.03$ is negligible for the present study (see additional discussion in supplementary material sections 1.4 and 1.7).

Next, to investigate the dependence of the polarization transfer rate on different dressed states, we park the Rabi frequencies at the HH matching conditions and vary the spin-lock (SL) duration. First, without any driving field applied on NV$_B$, the NV$_A$ SL signal is measured by sweeping the SL duration time as a reference. Under the SHH matching condition, driven by $\Omega_+^B = 7.56$ MHz, NV$_A$ SL coherence is drastically lost at a rate of $119 \pm 10$ kHz, which is extracted from a fit to the data (Fig. 4d). For $\Omega^A \gg \nu_{dip}$, the calculated effective dipolar coupling strength from Eqn. (2) is $\nu_{eff} \approx \nu_{dip}/2 \approx 130$ kHz, which agrees well with our measurement. Under the DHH matching condition, driven by $\Omega_+^B = 9.59$ MHz and $\Omega_-^B = 4.13$ MHz, NV$_A$ SL coherence is lost with a reduced rate of $73 \pm 10$ kHz compared to the SHH condition. This result indicates a reduced effective coupling between NV$_A$ and NV$_B$; the calculated effective coupling strength is $\nu_{eff} \approx 89$ kHz, which is roughly consistent with our measurement. Note that the measured polarization transfer rates are somewhat lower than the calculated values; also polarization return back to NV$_A$ does not happen with full contrast. We suspect such non-ideal behavior is due to imperfect HH matching conditions resulting from coplanar waveguide heating and drift of the external bias magnetic field during the measurements.

**E. Simulation of interaction dynamics in a spin ensemble**

We simulate strongly-coupled NV spin ensemble dynamics for the doubly dressed state, using a semi-classical model. The doubly dressed state scheme induces a homogeneity of spin interaction strengths, which can enhance the fidelity of generating many-spin entangled states[29]. Here, we assume a 50 ppm NV concentration with no other defect spin species present; implying a mean distance between NV spins of ~5 nm, which is in the strong coupling regime for a Carr-Purcell-Meiboom-Gill (CPMG) pulse enhanced decoherence rate $1/T_2 < 0.2$ MHz[30]. In reality, this 50 ppm NV diamond sample also carries significant concentrations of unwanted bath electronic defect spins causing short NV $T_2 < 1$ μs. However, the doubly dressed scheme, realized via the SL sequence, largely decouples the NV spins from the bath spins[1], facilitating observation of NV-NV spin interaction dynamics. Note that NV many-body interactions can lead to inhomogeneous detuning of individual NV centres from the globally applied Rabi driving field's central frequency, which can affect (reduce) the effective dipolar coupling strength. We find that this detuning effect is limited and dominated by the coupling strength distribution of the NV spin ensemble (see supplementary material section 1.4). One way to understand the ensemble spin dynamics is to adopt a spin bath spectral density model, parameterized with $\Delta$, the spin to bath coupling, and $R_{dd}$, the pairwise bath spin-spin coupling[26]. In the simulation, NV ensemble spins are simplified into a collection of two-spin pairs, similar to the approximation applied in a second order cluster correlation expansion (CCE) calculation[31,32] (see supplementary material section 3). Including higher-order spin correlations in the simulation could, in principle, better estimate the ensemble spin dynamics. However, due to the complexity of the problem, only pairwise clusters are considered here. Microscopic understanding of clusters of three or more electronic spins has been proposed using the CCE method[33]. However, the coherence function between spin pairs sometimes becomes 0, which causes a divergence in triple or higher spin correlations, making calculations very challenging. We note that an approximate treatment of higher-order spin correlations is as a mean-field detuning induced by long-range interactions on each spin pair[33]. With the doubly dressed state scheme, this mean-field detuning can be easily adjusted, which may offer a promising avenue to study higher-order spin correlation dynamics in future work.

We use a central spin model and extract an overall effective dipolar interaction $\Delta^2 = \sum_k v_{eff,k}^2$ between the central NV and off-axis NV bath spins[26]. We only consider interactions with off-axis NV spins because for spins with the same axis, spin exchange is highly suppressed in the dressed state due to the NV's spin-1 nature and the energy conservation in the rotating frame[1]. Different lattice configurations are simulated and contribute to a statistical distribution of Δ values. The effect of the doubly dressed state scheme is included by adding driving field terms to the spin pair Hamiltonian. The effective dipolar interaction strength distribution follows a probability density function (PDF); and by assessing the resulting PDF peak position and full width half maximum (FWHM) values, we can estimate the overall spin interaction dynamics. With single transition driving of Rabi frequency $\Omega_+ = 10$ MHz ($|0\rangle \leftrightarrow |+1\rangle$), both the PDF peak and FWHM are reduced by almost half compared to that for the not driven (ND) case (Fig. 5a). This trend continues as an additional second driving field $\Omega_-$ ($|0\rangle \leftrightarrow |-1\rangle$) is introduced, pushing Δ from the strongly coupled to weakly coupled regime. With $\Omega_+ = 10$ MHz and $\Omega_- = 8$ MHz, Δ is peaked at 48 ± 10 kHz, compared to the expected effective coupling of 43 kHz deduced from the Eqn. (2) using a bare dipolar coupling of 390 kHz extracted from the ND case (Fig. 5b). Next, we select pairs of off-axis NVs with the strongest coupling over the ensemble of spins and simulate the statistical distribution of pairwise dipole interaction strengths, $max(v_{ij}) = R_{dd}$. Among the four different NV crystalline axis classes, one class is fixed with no driving and the three other off-axis classes are driven. With single transition driving of $\Omega_+ = 10$ MHz, both the PDF peak and FWHM are reduced almost in half (Fig. 5c); and the trend continues as a second driving field $\Omega_-$ is applied. For example, the FWHM for doubly dressed states is 80 ± 10 kHz with $\Omega_+ = 10$ MHz, $\Omega_- = 8$ MHz; whereas the effective dipolar coupling value from Eqn. (2) is $v_{eff} \approx 107$ kHz for an ND FWHM of 973 kHz, implying more uniformity in collective coupling strengths (Fig. 5d). For both Δ and $R_{dd}$, we also confirm the homogenization of coupling strengths through the convergence of FWHM/Peak values for doubly dressed states at higher double transition Rabi frequencies (Fig. 5a,c). From the simulation results, the doubly dressed state scheme is expected to both suppress interactions between the central NV spin and off-axis NV bath spins, and reduce variation of spin-spin coupling strengths, making the ensemble spin system more uniform for coherent manipulation. Such a homogenized interaction distribution, combined with the extended long spin coherence time due to the SL measurement, may improve the fidelity of non-classical state generation and extend spin entanglement lifetimes (see supplementary material section 4).

**Summary and outlook**

To summarize, we demonstrate experimentally the use of dressed-state techniques to control the effective dipolar interaction in a strongly-coupled, solid-state electronic spin-1 system. Using a strongly-coupled pair of nitrogen vacancy (NV) centres in diamond as a model system, with $\nu_{dip}$ being the bare NV-NV dipolar coupling strength, we induce Rabi driving between different ground spin state sub-levels ($|0\rangle \leftrightarrow |\pm 1\rangle$) and employ a doubly dressed-state to tune the effective coupling strength between $-\nu_{dip}/2 < \nu_{eff} < +\nu_{dip}/2$, which is spectroscopically observed via Ramsey measurements. Other pulse schemes[5,34] to manipulate or suppress effective couplings are comparatively more complicated, with the duration and fidelity of the engineered Hamiltonian typically limited by pulse errors. In contrast, the doubly dressed state scheme provides a robust method to tune the effective coupling dynamics in a qutrit system once the driving strength is larger than the bare dipolar coupling strength $\nu_{dip}$. Furthermore, since the effective coupling strength depends only on the ratio of driving field amplitudes, systematic effects due to inhomogeneous Rabi driving amplitudes over the spin ensemble are highly suppressed. In principle, the doubly dressed state scheme should be applicable to a larger number of spin clusters, particularly when each NV spin can be selectively controlled. We envision tuning the higher order interaction dynamics by utilizing different dressed states for different NV spins. This method could be used to control the order parameter in a disordered spin system to study the transition of non-equilibrium phases[1]. Furthermore, reducing the local distribution of spin couplings could be used to increase fidelity in the generation of collective non-classical states. For example, creating an emergent Greenberger–Horne–Zeilinger (GHZ) state via quantum domino dynamics[35] in an Ising spin chain largely depends on the interaction uniformity[29,36]. Also, better fidelity generation of a many-spin Schrodinger cat state could be a valuable resource for enhanced quantum sensing or quantum information applications.

**Methods:**

**Experimental setup**

Measurements are conducted using a home-built NV-diamond confocal microscope setup. An acousto-optic modulator (Isomet Corporation) allows time-gating of a 400 mW, 532 nm diode-pumped solid-state laser (Changchun New Industries). The laser beam is coupled to a single-mode fiber, then delivered to an oil-immersion objective (100x, 1.3 NA, Nikon CFI Plan Fluor), and focused onto a diamond sample. The diamond sample is fixed on a three-axis motorized stage (Micos GmbH) for precise position control. NV red fluorescence (FL) is collected back through the same objective, then passes through a dichroic filter (Semrock LP02-633RS-25). A pinhole (diameter 75 $\mu$m) is used with a f = 150 mm telescope to spatially filter the FL signal, which is detected with a silicon avalanche photodetector (Perkin Elmer SPCM-ARQH-12). A signal generator (SG, Agilent E4428C) provides the carrier microwave signal. A 1G/s rate arbitrary waveform generator (AWG, Tektronix AWG 5014C) phase and amplitude modulates the carrier signal via an IQ mixer (Marki IQ 1545 LMP). Two resulting output microwave signals are amplified (Mini-circuits ZHL-16W-43-S+), combined, and sent through a gold coplanar waveguide fabricated on a glass cover-slip by photo-lithography and mounted directly on the diamond sample in order to manipulate the NV spins.

**Diamond sample**

Creation of a strongly-coupled NV pair is done by molecular ion implantation (Innovion corp), with a 6 keV energy $^{+28}$N molecular beam and an implantation dosage of $1\times10^9$/cm$^2$ applied to a diamond substrate. The diamond is CVD grown, $^{12}$C isotopically purified to 99.99%, and has dimensions of 2 mm × 2 mm × 0.5 mm. After ion implantation, the diamond is annealed at 800°C for 8 hours and at 1000°C for 10 hours to enhance conversion of N to NV and optimize NV optical and spin properties. Statistical measurement of NV FL intensity reveals that ∼ 5% of the NVs consist of proximal (few nm separation) NV pairs, limited by the conversion of N to NV. 6 keV ion implantation creates NV pairs with an average separation of ∼ 6 nm[1]; this corresponds to ∼ 0.2 MHz magnetic dipolar coupling strength between the two NV spins. Using the DEER measurement technique described in the main text, the NV pair coupling strength in the diamond sample ranges from about 0.050 MHz to 0.8 MHz, for several strongly-coupled NV pairs studied. The specific NV pair used in the reported measurements has spin lifetimes given in Table 1.

|       | NV$_A$         | NV$_B$         |
|-------|----------------|----------------|
| $T_1$ | $4.3 \pm 0.2$ ms | $4.3 \pm 0.2$ ms |
| $T_2$ | $49.7 \pm 4.5 \mu s$ | $13.3 \pm 1.2 \mu s$ |
| $T_2^*$ | $7.2 \pm 0.5 \mu s$ | $2.1 \pm 0.2 \mu s$ |

Table 1: Measured spin lifetimes for the two strongly-coupled off-axis NVs used in the reported experiments.

**Calibrations**

We applied a bias magnetic field of approximately 45 G, aligned with one of the NV spins, to ensure that the detuning between the two NV spins ($\Delta > 60$ MHz) was sufficiently large to suppress the crosstalk error effect. The driven Rabi frequency was carefully selected to satisfy the conditions $\Delta \gg \Omega_{\pm} \gg \nu_{dip}, h$, where $h$ denotes the hyperfine constant, so that the interaction dynamics could follow the simple picture provided in Eqn. (1). Calibration of the Rabi frequencies for each NV spin was achieved by initializing one of the NV spins and fitting the Rabi nutation measurement to a simple Hamiltonian model with a $^{14}$N hyperfine interaction. To mitigate the temperature rise issue during the continuous Rabi driving, we set the duty cycle of the microwave signal to 10%.

**Acknowledgments**

This material is based upon work supported by, or in part by, the US Army Research Laboratory and the U.S. Army Research Office under contract/grant numbers W911NF1510548 and W911NF1110400; the NSF Electronics, Photonics, and Magnetic Devices program under Grant No. ECCS-1408075; the NSF Physics of Living Systems program under Grant No. PHY-1504610; the Integrated NSF Support Promoting Interdisciplinary Research and Education program under Grant No. EAR1647504; the Army Research Laboratory MAQP program under Contract No. W911NF-19-2-0181; and the University of Maryland Quantum Technology Center. This work was performed in part at the Center for Nanoscale Systems, a member of the National Nanotechnology Coordinated Infrastructure Network, which is supported by the NSF under Grant No. 1541959. JL was supported by an ILJU Graduate Fellowship and the KIST open

16research program (2E31531). MT is supported by JSPS fellowship (JSPS KAKENHI Grant No. 20J01757). We thank Keigo Arai, Huiliang Zhang, and Hosung Seo for helpful discussions.

**Author contribution**

J.L. conceived the idea and R.L.W. supervised the project. J.L., M.T., and A.X. developed analytical and numerical simulations. J.L. performed the measurements, and analysed the data. All authors discussed the results and participated in writing the manuscript.

**Competing Financial Interests**

The authors declare no competing financial interests.

**Data availability**

The data supporting the findings of this study are available from the corresponding author upon reasonable request.

**Code availability**

The codes used in this study are available from the corresponding author upon reasonable request.

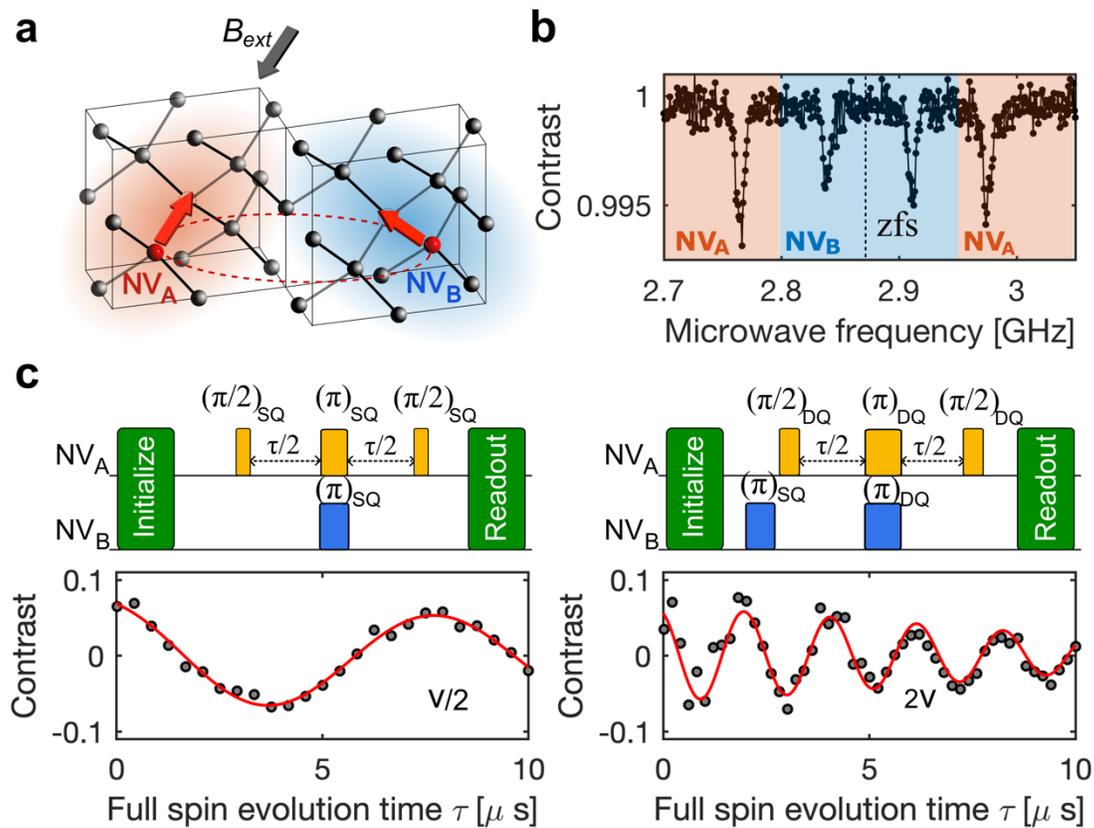

**Figure 1. Characterization of two NV qutrit spin system. a,** Schematic of two strongly-coupled NVs inside the diamond lattice, separated by ~10 nm. External bias magnetic field $B_{ext}$ is aligned with the quantization axis of the sensor spin NV$_A$ (red); the control spin NV$_B$ (blue) has a different quantization axis. **b,** Optically detected magnetic resonance (ODMR) measurement of NV pair system. With different Zeeman splittings due to different $B_{ext}$ field projections on each NV

quantization axis, NV$_A$ and NV$_B$ are resolved in the frequency domain. **c,** Double electron-electron resonance (DEER) measurement pulse sequences and measurement results. NV$_A$ is used as a sensor spin in both the single quantum (SQ, left) and double quantum (DQ, right) basis. NV$_B$ is used as a control spin, $\pi$ flipped from $m_s = 0$ to $m_s = +1$ (left) and double $\pi$ flipped from $m_s = -1$ to $m_s = +1$ (right). From the resulting modulation frequencies of the SQ and DQ DEER measurements, we extract the NV$_A$-NV$_B$ dipolar coupling strength of $\nu_{dip} = 0.250 \pm 0.015$ MHz.

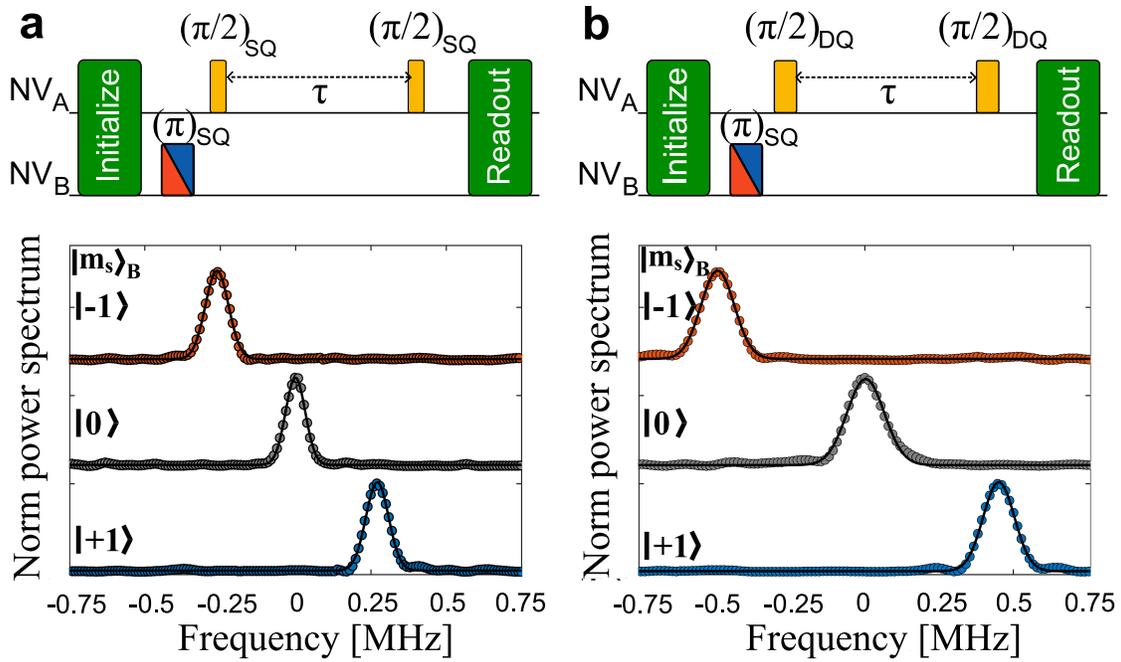

**Figure 2. Ramsey spectroscopy for sensing dipolar coupling strength of two NV qutrit system. a,** Single quantum (SQ) Ramsey spectroscopy pulse sequence varies the evolution time $\tau$ of the sensor spin NV$_A$ with the control spin NV$_B$ initialized to $|0\rangle$ (gray), $|+1\rangle$ (blue) and $|-1\rangle$ (red) states. Relative shifts in the peak of the power spectrum of the Ramsey signal as a function of the state of NV$_B$ give the dipolar coupling strength $\nu_{dip} = 0.26 \pm 0.02$ MHz. **b,** Repeat of the same measurement for the double quantum (DQ) basis of NV$_A$. Due to doubling of the effective magnetic moment of spin NV$_A$ in the DQ basis, twice larger shifts are observed in the Ramsey signal power spectrum, yielding $2\nu_{dip} = 0.52 \pm 0.02$ MHz.

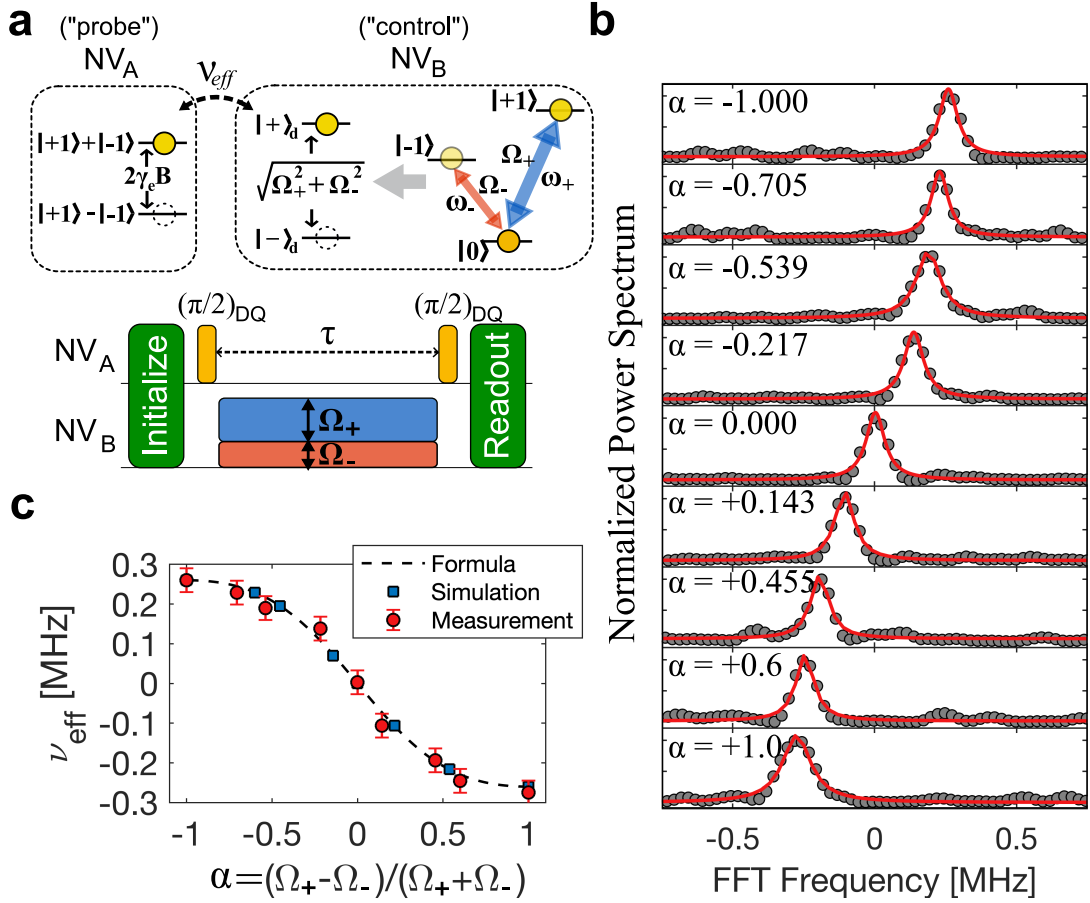

**Figure 3. Tuning effective coupling strength of two NV qutrit system via doubly dressed state. a,** Left inset box shows the double quantum state energy level of the sensor spin NV$_A$. Right inset box shows the ground state energy level of the control spin NV$_B$. Microwave control fields at resonance frequencies $\omega_+$, (between $|0\rangle$ and $|+1\rangle$ states) and $\omega_-$ (between $|0\rangle$ and $|-1\rangle$ states) are driven with Rabi frequencies $\Omega_\pm$, which results in an effective two-level doubly dressed state $|+\rangle_d$, $|-\rangle_d$ shown in the right inset box. Below is the pulse sequence for DQ Ramsey spectroscopy on the sensor spin NV$_A$ with NV$_B$ driven by microwave control fields at $\Omega_\pm$. **b,** Normalized power spectrum of the Ramsey signal for NV$_A$. Gray dots are measurement and red solid lines are Lorentzian fits to the data. Dressed-state control parameter $\alpha = \frac{(\Omega_+ - \Omega_-)}{(\Omega_+ + \Omega_-)}$. For NV$_B$ driven on only a single transition, i.e., for $\alpha = \pm 1$, a modulation peak is observed at the bare dipolar coupling strength between the two NVs ($\pm \nu_{dip}$) because phase is accumulated in NV$_A$'s DQ basis. For NV$_B$ driven on both transitions with the same Rabi frequency, i.e., for $\alpha = 0$, a peak appears at FFT frequency = 0. As $\alpha$ is swept from $-1$ to $+1$ with $\Omega_\pm >$ 2 MHz $> \nu_{dip}$, the modulation peak

continuously shifts from $\nu_{eff} = +\nu_{dip}$ to $\nu_{eff} = -\nu_{dip}$. Here, the twice larger variation in measured $\nu_{eff}$, compared to Eqn. (2), is due to the Ramsey measured in DQ basis. **c,** Variation of $\nu_{eff}$ with dressed-state control parameter $\alpha$. Red dots are measurements, blue dots are from a numerical simulation, and black dashed line is from DQ corrected Eqn. (2). Simulation is from numerically solving Eqn. (1).

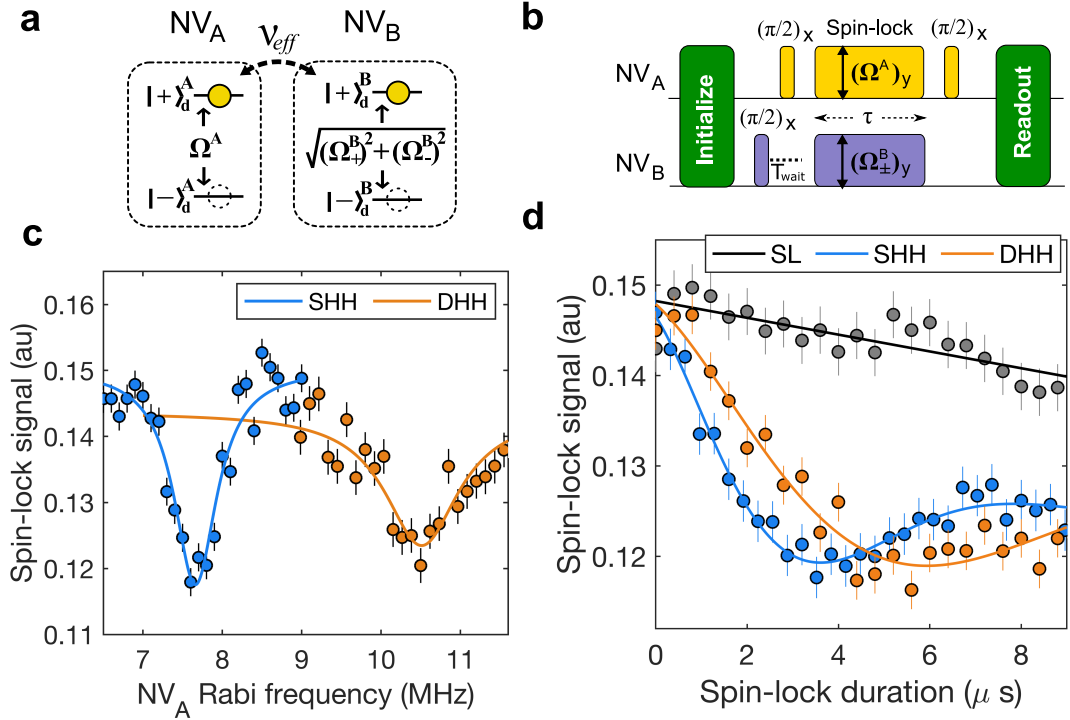

**Figure 4. Spin polarization transfer measurement under singly and doubly dressed state Hartmann-Hahn conditions (SHH and DHH). a,** Effective spin-1/2 model for cross-polarization between singly dressed $NV_A$ and doubly dressed $NV_B$. **b,** Polarization transfer pulse sequence. $NV_B$ is first prepared in a $|0\rangle, |+1\rangle$ mixed state and then Rabi driven to dressed states, while $NV_A$ is spin-locked along the y-axis. **c,** $NV_A$ spin-lock (SL) coherence signal measurement. While sweeping the $NV_A$ Rabi frequency. $NV_B$ is driven to SHH or DHH by controlling the $NV_B$ Rabi frequency $\Omega_\pm^B$. Once either matching condition is satisfied, $\Omega^A = \sqrt{(\Omega_+^B)^2 + (\Omega_-^B)^2}$, polarization from $NV_A$ is lost, as shown as dips in the SL signal. SHH is when $NV_B$ is singly driven, and DHH is when $NV_B$ is doubly driven. **d,** Polarization transfer dynamics measured via the $NV_A$ SL

coherence signal over the duration of $NV_A$ spin-lock driving. $NV_A$ Rabi frequencies are fixed at the SHH or DHH matching frequencies with $NV_B$.

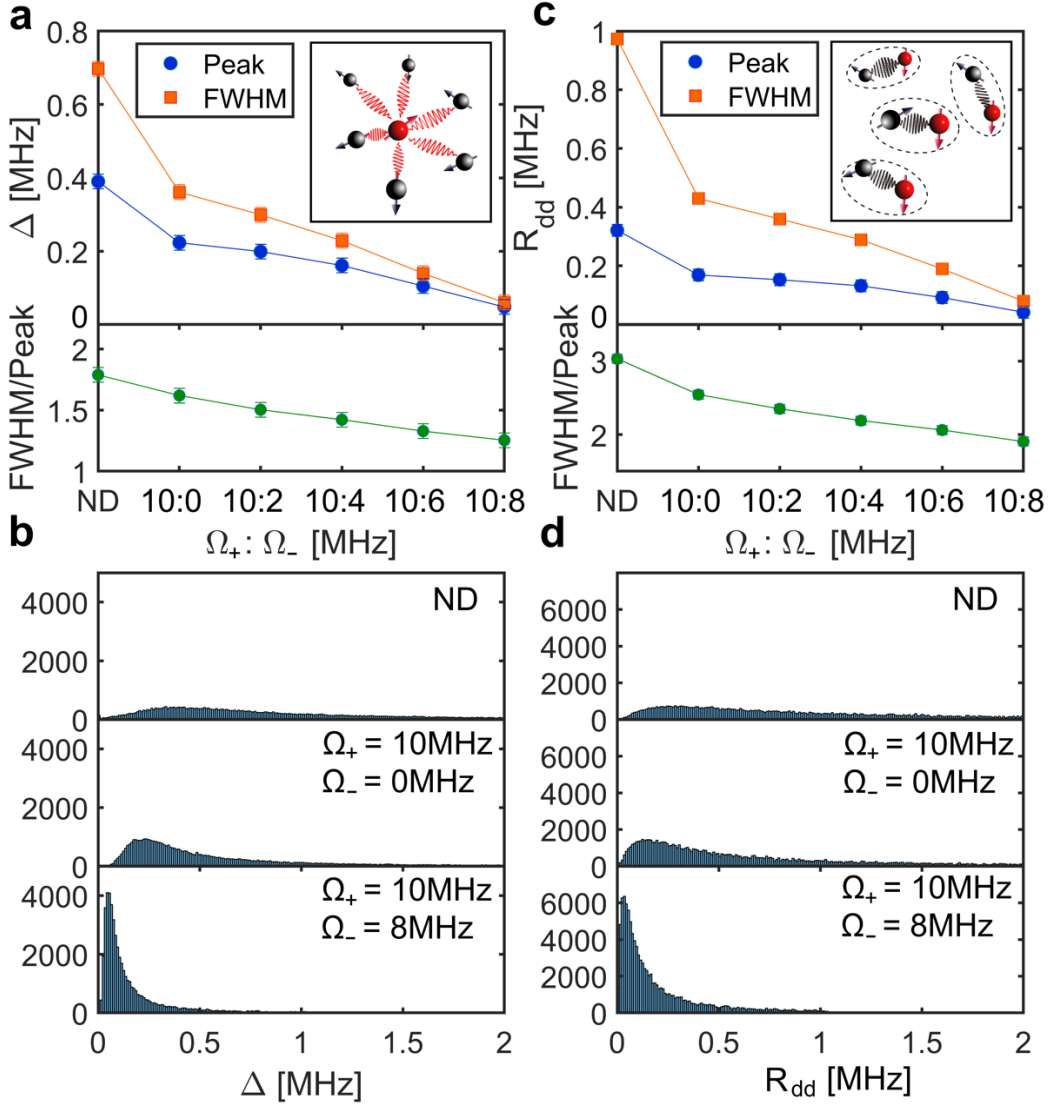

**Figure 5. Semi-classical simulation of effective dipolar couplings for an NV spin ensemble in the doubly dressed state. a,** Using a central spin model (inset), the effective coupling strength $\Delta$ between the central NV and the off-axis NV spin bath is calculated for multiple lattice configurations by varying the double transition driving field Rabi frequencies. The resulting $\Delta$ probability peak and full width at half maximum (FWHM) are then extracted and displayed (upper panel). Lower panel shows a $\Delta$ trend towards greater homogenization at higher double transition Rabi frequencies, as illustrated in the ratio of the FWHM/Peak for the doubly dressed states. **b,**

Probability distribution function (PDF) for $\Delta$ when off-axis bath NV spins are not driven (ND); singly driven $\Omega_+ = 10$ MHz, $\Omega_- = 0$; and doubly driven with $\Omega_+ = 10$ MHz, $\Omega_- = 8$ MHz. **c,** Upper panel: Extracted PDF properties for pairwise flip-flop rates $R_{dd}$ by varying $\Omega_\pm$. As shown in the inset figure, among 4 different NV crystalline axis classes, one class (red) is fixed with no driving and 3 other off axis classes (black) are driven. Lower panel shows homogenization trend of $R_{dd}$'s FWHM/Peak (green) for the doubly dressed states. **d,** PDF for $R_{dd}$ when pairwise off-axis NV spins are not driven (ND); singly driven with $\Omega_+ = 10$ MHz, $\Omega_- = 0$; and doubly driven with $\Omega_+ = 10$ MHz, $\Omega_- = 8$ MHz.